\documentclass[sigconf]{acmart}
\usepackage{makecell,multirow,color,booktabs,balance}
\AtBeginDocument{%
  }

\setcopyright{acmcopyright}
\copyrightyear{2022}
\acmYear{2022}
\acmDOI{10.1145/3552466.3556528}

\acmConference[DDAM '22]{Proceedings of the 1st International Workshop on Deepfake Detection for Audio Multimedia}{Oct 10--14, 2022}{Lisbon, Portugal}
\acmBooktitle{Proceedings of the 1st International Workshop on Deepfake Detection for Audio Multimedia (DDAM '22), October 14, 2022, Lisboa, Portugal}
\acmPrice{15.00}
\acmISBN{978-1-4503-9496-3/22/10}

\begin{document}

\title{Deepfake Detection System for the ADD Challenge Track 3.2 Based on Score Fusion}

\author{Yuxiang Zhang}
\email{zhangyuxiang@hccl.ioa.ac.cn}
\affiliation{%
	\institution{Institute of Acoustics, Chinese Academy of Sciences}
	\streetaddress{No. 21 North 4th Ring Road}
	\city{Haidian District}
	\state{Beijing}
	\country{China}
	\postcode{100190}
}
\affiliation{%
	\institution{University of Chinese Academy of Sciences}
	\streetaddress{No.19(A) Yuquan Road}
	\city{Shijingshan District}
	\state{Beijing}
	\country{China}
	\postcode{100049}
}

\author{Jingze Lu}
\email{lujingze@hccl.ioa.ac.cn}
\affiliation{%
	\institution{Institute of Acoustics, Chinese Academy of Sciences}
	\streetaddress{No. 21 North 4th Ring Road}
	\city{Haidian District}
	\state{Beijing}
	\country{China}
	\postcode{100190}
}
\affiliation{%
	\institution{University of Chinese Academy of Sciences}
	\streetaddress{No.19(A) Yuquan Road}
	\city{Shijingshan District}
	\state{Beijing}
	\country{China}
	\postcode{100049}
}

\author{Xingming Wang}
\email{xingming.wang@dukekunshan.edu.cn}
\affiliation{%
	\institution{Data Science Research Center, Duke Kunshan University}
	\streetaddress{No. 8 Duke Avenue}
	\city{Kunshan}
	\state{Jiangsu Province}
	\country{China}
	\postcode{215316}
}
\affiliation{%
	\institution{School of Computer Science, Wuhan University}
	\city{Wuhan}
	\state{Hubei Province}
	\country{China}
	\postcode{430070}
}


\author{Zhuo Li}
\email{lizhuo@hccl.ioa.ac.cn}
\affiliation{%
	\institution{Institute of Acoustics, Chinese Academy of Sciences}
	\streetaddress{No. 21 North 4th Ring Road}
	\city{Haidian District}
	\state{Beijing}
	\country{China}
	\postcode{100190}
}
\affiliation{%
	\institution{University of Chinese Academy of Sciences}
	\streetaddress{No.19(A) Yuquan Road}
	\city{Shijingshan District}
	\state{Beijing}
	\country{China}
	\postcode{100049}
}

\author{Runqiu Xiao}
\email{xiaorunqiu@hccl.ioa.ac.cn}
\affiliation{%
	\institution{Institute of Acoustics, Chinese Academy of Sciences}
	\streetaddress{No. 21 North 4th Ring Road}
	\city{Haidian District}
	\state{Beijing}
	\country{China}
	\postcode{100190}
}
\affiliation{%
	\institution{University of Chinese Academy of Sciences}
	\streetaddress{No.19(A) Yuquan Road}
	\city{Shijingshan District}
	\state{Beijing}
	\country{China}
	\postcode{100049}
}

\author{Wenchao Wang}
\email{wangwenchao@hccl.ioa.ac.cn}
\affiliation{%
	\institution{Institute of Acoustics, Chinese Academy of Sciences}
	\streetaddress{No. 21 North 4th Ring Road}
	\city{Haidian District}
	\state{Beijing}
	\country{China}
	\postcode{100190}
}

\author{Ming Li}
\email{ming.li369@dukekunshan.edu.cn}
\affiliation{%
	  \institution{Data Science Research Center, Duke Kunshan University}
	  \streetaddress{No. 8 Duke Avenue}
	  \city{Kunshan}
	  \state{Jiangsu Province}
	  \country{China}
	  \postcode{215316}
	}
\affiliation{%
	\institution{School of Computer Science, Wuhan University}
	\city{Wuhan}
	\state{Hubei Province}
	\country{China}
	\postcode{430070}
}

\author{Pengyuan Zhang}
\authornote{Corresponding author.}
\email{zhangpengyuan@hccl.ioa.ac.cn}
\affiliation{%
	\institution{Institute of Acoustics, Chinese Academy of Sciences}
	\streetaddress{No. 21 North 4th Ring Road}
	\city{Haidian District}
	\state{Beijing}
	\country{China}
	\postcode{100190}
}
\affiliation{%
	\institution{University of Chinese Academy of Sciences}
	\streetaddress{No.19(A) Yuquan Road}
	\city{Shijingshan District}
	\state{Beijing}
	\country{China}
	\postcode{100049}
}
\renewcommand{\shortauthors}{Yuxiang Zhang et al.}
\begin{abstract}
  This paper describes the deepfake audio detection system submitted to the Audio Deep Synthesis Detection (ADD) Challenge Track 3.2 and gives an analysis of score fusion. The proposed system is a score-level fusion of several light convolutional neural network (LCNN) based models. Various front-ends are used as input features, including low-frequency short-time Fourier transform and Constant Q transform. Due to the complex noise and rich synthesis algorithms, it is difficult to obtain the desired performance using the training set directly. Online data augmentation methods effectively improve the robustness of fake audio detection systems. In particular, the reasons for the poor improvement of score fusion are explored through visualization of the score distributions and comparison with score distribution on another dataset. The overfitting of the model to the training set leads to extreme values of the scores and low correlation of the score distributions, which makes score fusion difficult. Fusion with partially fake audio detection system improves system performance further. The submission on track 3.2 obtained the weighted equal error rate (WEER) of 11.04\%, which is one of the best performing systems in the challenge.
\end{abstract}



\maketitle

\section{Introduction}
\label{sec:intro}
Over the past few years, audio deepfake technology, including speech synthesis and voice conversion (VC) technologies have made great strides with the development of deep learning. State-of-the-art (SOTA) algorithms can generate high-quality fake audio that is difficult to distinguish even for human\cite{muller2021human}. With the development of technologies such as virtual reality and metaverse, the Internet is increasingly flooded with audio and video information. As a result, fake audio generated with deep synthesis algorithms may be used maliciously, posing a great threat to property security and social stability.

Attacking an automatic speaker verification (ASV) system with spoof speech generated by text-to-speech (TTS) \cite{wang2017tacotron, kong2020hifi} and VC \cite{kaneko2018cyclegan} is defined as the logical access (LA) attack. To improve the performance of the LA spoof speech attack countermeasures, the ASVspoof challenge \cite{wu2015asvspoof, Kinnunen2017, todisco2019asvspoof, yamagishi21_asvspoof} has played an important leading role in facilitating research on spoof speech detection. The challenge provides standard datasets containing various TTS and VC algorithms, as well as evaluation metrics. The ASVspoof 2021 had proposed Deepfake (DF) task, containing multiple TTS and VC algorithms, and undergoes multiple compression codecs. This task wants to verify the performance and robustness of the countermeasures with data in real scenarios.

The Audio Deep Synthesis Detection Challenge (ADD 2022)\cite{yi2022add} goes a step further. Considering the realistic situation with noise, partially fake audio and fast updates of TTS and VC algorithms, the ADD 2022 Challenge proposes three tracks. Track 1 is the low-quality fake audio detection task. The real and fake audio in the test set has various noise and background music. Track 2 is the partially fake audio detection task. The test set includes partially fake utterances generated by manipulating the real speech with real or synthetic utterences. Track 3 of the ADD 2022 Challenge is audio fake game, of which track 3.1 is the generation task and track 3.2 is the detection task. There are two rounds of evaluations in track 3.2. The first test set is similar to that of track 1, while the test set of the second round consists of the first test set and a portion of the generated speech submitted by participants in the generation task. Thus, the confrontation between fake audio generation and detection is simulated, and the robustness of the generation and detection systems are effectively evaluated. And score fusion of different single countermeasures is a common method to improve the robustness of the primary systems.

The proposed countermeasures based on light convolutional neural network (LCNN) model \cite{tomilov21_asvspoof}, additive margin Softmax (AM-Softmax) Loss \cite{wang2018additive} and Center Loss \cite{wen2016discriminative}. Different features, including various short-time Fourier transform (STFT) and Constant Q transform (CQT) \cite{todisco2016new} spectrograms are used as front-ends. The system description also explores the effects of online data augmentation methods, including audio normalization, noise and reverberation addition, and speed perturbation. The data augmentation methods effectively improve the robustness of the fake audio detection system.

The primary systems are obtained by the score fusion of multiple single systems. However, the performance improvement of system score fusion in the track 3.2 is not obvious compared to the ASVspoof 2019 challenge \cite{lavrentyeva2019stc, lai2019assert, chettri2019ensemble}. The number of fusion scores, how relevant these scores are, and the fusion technique used determine the additional benefit of fusion \cite{ulery2006studies}. Therefore, through the visualization of the score distributions, the reasons for the insignificant improvement of the score fusion are analyzed. By observing the distribution of scores, two reasons for the failure of score fusion can be obtained. On the one hand, it is difficult to choose the weights of the score fusion because different systems are very confident to give different judgments to an audio due to the overfitting to the training set. On the other hand, system fusion cannot correct misjudged samples due to the poor correlation between the scores of different systems.

The main contribution of this work are summarized as follows:
\begin{itemize}
	\item \textbf{Proposed a fake audio detection system for the ADD 2022 challenge track 3.2.} The proposed LCNN-based score fusion system achieves Equal Error Rates (EERs) of 9.59\%, 11.96\%, and 11.04\% in the first round, second round, and final ranking, respectively, obtaining competitive results in the challenge through a simple approach.
	\item \textbf{The effect of audio normalization and other data augmentation methods are explored.} In addition to the common data augmentation methods such as noise addition, reverberation and speed perturbation, audio normalization has obvious effects on the adaptation set of the ADD challenge track 1. The effects of different data augmentation methods on the adaptation set are also compared.
	\item \textbf{The reasons for the poor fusion effect of system scores are analyzed.} Last but not least, the EER obtained by score fusion in the challenge is less than 10\% lower than the best single system, which is worse than the fusion effect on the ASVspoof 2019 dataset. Therefore, the reasons for the failure of score fusion is analyzed by visualizing the scores of different systems and and conducting a comparative analysis, the reasons for the failure of score fusion are explored.
\end{itemize}

\section{Related Work}
In this section, previous work in the field of fake audio detection is summarized. First, the commonly used features and models for anti-spoofing are summarized. After that, data augmentation methods for fake audio detection are summarized. Finally, the application of score fusion in fake audio detection is summarized.

\subsection{Features and models for fake audio detection}
Fake aduio detection systems usually consist of feature extraction and classifiers. A class of commonly used classifiers are based on Convolutional Neural Networks (CNN) such as Residual Neural Networks (ResNet)\cite{lai2019assert, chen21b_asvspoof, li2021replay} and Light Convolutional Neural Networks (LCNN)\cite{lavrentyeva2019stc, wang21fa_interspeech, tomilov21_asvspoof}. The input features of CNN based classifiers are two-dimensional features, including short-time Fourier transform (STFT) spectrograms, constant-Q transform (CQT) spectrograms, mel-scale transforms (MSTFT), and cepstral features such as constant-Q cepstral coefficients (CQCC) \cite{ todisco2017constant} and Linear Frequency Cepstral Coefficients (LFCC) \cite{sahidullah15_interspeech}. Recently, network models with raw audio as input have also appeared, such as RawNet2 \cite{9414234, tak21_asvspoof} and graph network based countermeasure \cite{9747766}. Network architecture search is also used for spoofing speech detection \cite{ge21c_interspeech, ge21_asvspoof}. On the ASVspoof 2019 LA database \cite{wang2020asvspoof}, the SOTA systems can even achieve an EER around 1\% \cite{9747766, tak21_asvspoof, zhang21da_interspeech}. The self-supervised front-end effectively improves the robustness of countermeasures against complex scenarios. Systems using wav2vec 2.0 \cite{baevski2020wav2vec} as the front-end achieve a huge performance boost in the ASVspoof 2021 DF task \cite{wang2021investigating}, ADD challenge track 1  \cite{9747768} and track 2 \cite{9747605} compared to systems using traditional features.

Despite showing excellent performance on specific datasets, current fake audio detection algorithms still suffer from some shortcomings. Deep learning-based countermeasures are not robust enough to unknown deepfake algorithms and out-of-distribution data \cite{wang2022estimating, muller2022does}. In addition, there is currently a lack of analysis of the physical meaning of algorithms and features, and the credibility of the algorithms is low.

\subsection{Data augmentation}
To improve the generalization of the system in real scenes, many data augmentation methods are applied to fake audio detection. In replay speech detection, the real speech can be adjusted by parametric voice reverberators and phase shifters \cite{cai2017countermeasures}. Data augmentation by speed perturbation is also used in replay detection \cite{cai2019dku}. For synthetic speech detection, noise and reverberation addition commonly used in other fields are also effective \cite{chen2020generalization, yan2022audio}. Channel compensation by compression coding can also be used for data augmentation \cite{das2021data}. Furthermore, in order to overcome the influence of channel variation, a variety of codec formats are used in data augmentation \cite{das21_asvspoof, chen21_asvspoof, chen21b_asvspoof}. Another online data augmentation method that works well in fake audio detection is the finite impulse response (FIR) filter \cite{tomilov21_asvspoof, 9747768}. Filtering the audio through a filter can mask different frequency bands, the effect is similar to SpecAug frequency masking \cite{park2019specaugment}. Based on a variety of convolutional and additive noises, RawBoost \cite{tak2022rawboost} models nuisance variability stemming such as encoding, transmission, microphones and amplifiers, as well as linear and nonlinear distortion. Due to the low robustness of fake audio detection systems, data augmentation is becoming an indispensable step in countermeasure training.

\subsection{Score fusion}
Score-level fusion is widely used in multimodal biometrics \cite{singh2008integrated, he2010performance, singh2019comprehensive}, as it provides the best trade-off in terms of information content and ease of fusion \cite{nandakumar2007likelihood}.  If it is assumed that the joint distribution of real and impostors is known, score fusion based on likelihood ratio is a commonly used score fusion method, including the product of likelihood ratio and logistic regression of likelihood ratio. When the true distribution is unavailable or unreliable, the scores can be normalized by transformation to false acceptance rates (FARs) before fusion, and fusion can be performed by the maximum, minimum or product of FARs. When the score distributions are unknown, simple linear fusion algorithms include average score fusion, weighted average score fusion, maximum score fusion, minimum score fusion, and medium number score fusion. Where multiple classifier score average or weighted average, maximum, minimum or the median is regarded as the final score. 

Score fusion is also widely used in the field of fake audio detection. In the ASVspoof 2015 \cite{xiao2015spoofing} and ASVspoof 2019 LA Challenges \cite{lavrentyeva2019stc, lai2019assert, chettri2019ensemble}, the performance of score fusion can be improved by more than 20\% compared to the best single system. However, combining matching scores is a challenging task, since scores for different matchers can be distance or similarity measures, may follow different probability distributions, may provide completely different accuracies and may be correlated \cite{nandakumar2007likelihood}. Therefore, the performance improvement of score fusion is insignificant on datasets such as ASVspoof 2021 where the data distribution differs significantly from that of the training set \cite{chen21b_asvspoof, wang21_asvspoof, kang21b_asvspoof}. However, the reason for the failure of score fusion has not been analyzed. Since the detection of unknown algorithms is the main application scenario of fake audio detection, and system fusion is a common method to improve performance of fake audio detection system. It is worth exploring how to effectively perform the fusion of fake audio detection systems

\section{Methods}
All of the proposed single systems are based on the LCNN architecture similar to \cite{tomilov21_asvspoof}, which is the best system in ASVspoof 2021 LA and DF tasks \cite{yamagishi21_asvspoof}. However, different features and loss functions are used in our submissions. The front-end, model architecture, loss function, and score fusion method of the proposed systems are all discussed separately in this section.

\subsection{Problem Statement and Overview}
Given an input audio, the problem is to detect whether it is a fake audio. Firstsly, multiple spectrogram features are extracted from the input audio. Then different features are used to train the LCNN model. Each model uses one of the loss functions in AM-Softmax Loss \cite{wang2018additive} or Center Loss \cite{wen2016discriminative}. As a result of screening through the EER for different models in the adaptation set for track 1, two single systems are selected in the first round and six single systems are selected in the second round. The normalized scores of the single systems are weighted summed based on the track 1 adaptation set, and the weights are obtained through logistic regression. The performance of score fusion depends upon the distributions of scores, as well as the correlation between scores. So the correlation of scores between single systems is analyzed by visualization.

\subsection{Front-End}
The main acoustic features used in our experiments are log power magnitude spectrograms, including STFT and CQT. The STFT features are extracted with librosa toolkit \cite{mcfee2015librosa}. According to \cite{zhang21da_interspeech}, the low frequency part below $4k$ Hz of the spectrogram is more robust. Similarly, in the ASVspoof 2021 Challenge, a good performance improvement has been achieved by decreasing the impact of spectral features of frequencies above $3.4k$ Hz through a finite impulse response low-pass filter \cite{tomilov21_asvspoof}. Thus only low-frequency part ($0-4$ kHz) of spectrograms are fed into neural networks. Features are utilized as the following configurations: 
\begin{itemize}
	\item STFT-1024: 1024 window length, 1024 number of FFT bins and 160 hop length. Only the low-frequency part is used.
	\item STFT-2048: 2048 window length, 2048 number of FFT bins and 160 hop length. The same low-frequency part is used.
	\item CQT: 49 bins per octave, minimum frequency is 15.6 and maximum frequency is 4000.
\end{itemize}

\subsection{LCNN classifier and loss function}
The proposed LCNN based system is similar to the LCNN based baseline system \cite{wang21fa_interspeech} and the system in \cite{tomilov21_asvspoof} in the ASVspoof 2021 challenge. The LCNN model consists of nine convolutional layers and two bidirectional long short-term memory (BLSTM) layers. The LCNN architecture is characterized by the use of Max-Feature-Map activation based on the Max-Out activation function \cite{goodfellow2013maxout}, which enables the selection of features that are critical for the task solving, reduces the number of parameters, and improves robustness of model. The detailed architecture is described in Table.\ref{lcnn}.

\begin{table}[ht]
	\centering
	\caption{The architecture of LCNN.}
	\scalebox{1}{
		\label{lcnn}
		\begin{tabular}{ l c r}
			\toprule
			\textbf{Type} & \textbf{Filter/Stride} & \textbf{Output} \\
			\midrule
			Conv1 & $5\times5$ / $1\times1$ & $257\times600\times64$\\
			MFM1 & - & $257\times600\times32$\\
			\midrule
			MaxPool1 & $2\times2$ / $2\times2$ & $128\times300\times32$\\
			\midrule
			Conv2 & $1\times1$ / $1\times1$ & $128\times300\times64$\\
			MFM2 & - & $128\times300\times32$\\
			BatchNorm1 & - & $128\times300\times32$\\
			Conv3 & $3\times3$ / $1\times1$ & $128\times300\times96$\\
			MFM3 & - & $128\times300\times48$\\
			\midrule
			MaxPool2 & $2\times2$ / $2\times2$ & $64\times150\times48$\\
			BatchNorm2 & - & $64\times150\times48$\\
			\midrule
			Conv4 & $1\times1$ / $1\times1$ & $64\times150\times96$\\
			MFM4 & - & $64\times150\times48$\\
			BatchNorm3 & - & $64\times150\times48$\\
			Conv5 & $3\times3$ / $1\times1$ & $64\times150\times128$\\
			MFM5 & - & $64\times150\times64$\\
			\midrule
			MaxPool3 & $2\times2$ / $2\times2$ & $32\times75\times64$\\
			\midrule
			Conv6 & $1\times1$ / $1\times1$ & $32\times75\times128$\\
			MFM6 & - & $32\times75\times64$\\
			BatchNorm4 & - & $32\times75\times64$\\
			Conv7 & $3\times3$ / $1\times1$ & $32\times75\times64$\\
			MFM7 & - & $32\times75\times32$\\
			BatchNorm5 & - & $32\times75\times32$\\
			Conv8 & $1\times1$ / $1\times1$ & $32\times75\times64$\\
			MFM8 & - & $32\times75\times32$\\
			BatchNorm6 & - & $32\times75\times32$\\
			Conv9 & $3\times3$ / $1\times1$ & $32\times75\times64$\\
			MFM9 & - & $32\times75\times32$\\
			\midrule
			MaxPool4 & $2\times2$ / $2\times2$ & $16\times37\times32$\\
			\midrule
			Flatten & - & $37\times512$\\
			BLSTM1 & - & $37\times512$\\
			BLSTM2 & - & $37\times512$\\
			MeanPool & - & $512$ \\
			\midrule
			FC1 & - & 128 or 512 \\
			Dropout & - & - \\
			FC2 & - & 2 \\
			\bottomrule
		\end{tabular}
	}
\end{table}

The features are extracted from the convolutional layers. After that they are pooled as a sequence by fed into the BLSTMs. Then, the inputs and outputs of BLSTMs are summed and averaged in the time domain. Finally the mean vector is reduced to a 512-dimentional or 128-dimensional embedding vector by fully connected layers and then projected to 2 dimensions. The 512-dimentional embedding vectors are used to calculate the AM-softmax Loss while the 128-dimential embedding vectors are used to calculate the Center Loss.

Both AM-Softmax Loss \cite{wang2018additive} and Center Loss \cite{wen2016discriminative} are first proposed for face recognition task. And their basic idea is to increase the inter-class distance and decrease intra-class distance. The AM-Softmax is an improvement to angular margin based softmax loss \cite{liu2017sphereface} that simplifies the calculation and speeds up convergence. The constraint of Center loss is relatively weaker, only the sum of squared distances from features to feature centers in a batch is constrained to be as small as possible, i.e., the intra-class distance is as small as possible. Therefore, it needs to be combined with the cross entropy loss and optimized jointly.

\subsection{System fusion}
The primary systems submitted to the challenge are fusion of single systems on the score level. The fusion weights are selected based on the performance as well as logistic regression fit of each single system on the track 1 adaptation set.

Due to the long duration of the audio in the Track 3.2 test set and the splicing traces found in it, the system used in track 2 to detect partially fake audio is utilized in the second round of track 3.2 evaluation.

\section{Experimental Setting and Implementation}

\subsection{Datasets}
All experiments presented are conducted on the ADD 2022 Challenge datasets. Only the training set is used to train all the systems. The development set is used for performance validation during training. Since the trained model have a very high correct rate on the adaptation set of track 3.2, this adaptation set is of little significance. Therefore, the adaptation set of track 1 is used for performance measurement and weight adjustment for system fusion. Since the audio in the training set is very clean, the training process converged quickly. While the test set is much more complex, containing noise and fake audio generated by other teams using new algorithms and post-processing, etc. Therefore, enhancing model generalization is the biggest challenge.

\subsection{Data augmentation and audio normalization}\label{dataaug}
Data augmentation is a common strategy used to increase the amount of training data. And it has been shown to be effective for training neural networks to make robust predictions. 

According to the evaluation plan, the track 3.2 test sets in both rounds are similar to track 1, which contain various noises and background music. Therefore, noise and reverberation from MUSAN\cite{musan2015} and RIRs\cite{ko2017study} datasets are added to the audio of training and development set online in a Kaldi \cite{povey2011kaldi} like manner. 

The audio in the training and development set is also coded and decoded by ffmpeg with different codec algorithms, including mp3, m4a, ogg, opus, alaw, $\mu$law and g.722.

By examining the data, the volume levels of the audio in the datasets are found to have large variances, with most of the audio being at lower volume. So the audio in training, developing and the test set is normalized with sv56 \cite{rec2005g}.

Combining the original data, codeced data, and normalized data yields a dataset nine times the number of the original dataset. A total of 120,000 entries are selected from it as training data. The development set is treated similarly, and randomly select 60,000 audio  from it.

\subsection{Chunk-size randomization}
Since the input data of each batch needs to be the same size during the training process, but the duration of audio always varies, the feature of each audio need to be fixed to a uniform length. The length of the audio is usually unified with truncation or padding. In speaker recognition and language recognition, variable-length training achieves better results than fixed-length training \cite{9036861}. Therefore, a similar variable-length training method is introduced into the detection systems. In the training process, for each batch, a random integer $N$ is first generated from the uniform distribution in the interval $(N_{min},N_{max})$ as the chunk-size of this batch. Later, when the data of this batch is loaded, the chunk-size of each feature is truncated or padded to $N$. In our system, the chunk-size of the feature is randomly selected in the interval of $(500,700)$.

\subsection{Training strategy}
The Adam optimizer is adopted with $\beta_1=0.9$, $\beta_2=0.98$, $\epsilon=10^{-9}$ and weigth decay $10^{-4}$. The learning rate is initialized as $0.0003$. As a scheduler, StepLR is used with step size of 10 epochs and coefficient 0.5.

The factor of Center Loss combined with the cross-entropy loss is $0.05$. The parameters of the AM-Softmax Loss are $s=20$ and $m=0.9$. 

The model with the lowest loss on the development set is selected as the final model for evaluation.

\subsection{Score Fusion}
If the range of scores changes, the fusion results will not be satisfactory \cite{kaya2017video}. Thus the scores of the different systems are max-min normalized:
$$
f(S)=\frac{S-min\{S_k\}}{max\{S_k\}-min\{S_k\}}
$$

\begin{table}[ht]
	\centering
	\caption{Final submission systems for Track 3.2}
	\label{weight}
	\vspace{2mm}
	\begin{tabular}{ l c c r}
		\toprule
		\multirow{2}*{\textbf{Feature}} & \multirow{2}*{\textbf{Loss}} & \multicolumn{2}{c}{\textbf{Weight}} \\
		\cline{3-4}
		& & 1st round & 2nd round \\
		\midrule
		STFT-1024 & AM-Softmax & 0.95 & 0.823 \\
		STFT-1024 & Center Loss & 0.05 & 0.118 \\
		STFT-1024-norm & AM-Softmax & - & -0.059 \\
		STFT-2048 & AM-Softmax & - & -0.118 \\
		STFT-2048 & Center Loss & - & 0.118 \\
		CQT & Center Loss & - & 0.118 \\
		\bottomrule
	\end{tabular}
\end{table}

The weights of the system score fusion are shown in Table \ref{weight}. The norm in features means that the audio is normalized before extracting features. The weights are selected based on the calculation results of the BOSARIS \cite{brummer2013bosaris} toolkit through logistic regression. The negative weights are obtained based on the fusion weights given by BOSARIS on the track 1 adaptation set. The weight of the best performing single system dominates, otherwise the fusion system will experience performance degradation.

In the second round, if the partially fake aduio detection system detects that the probability of the utterance being partially fake is greater than 77.5\%, the utterance is regarded as a partially fake audio. The scores of detected partially fake audio is set to the minimum of the scores of the fusion system.

\section{Results and Analysis}
The results of our single and fusion systems are expressed in terms of EER. The final ranking is based on the Weighted EER (WEER), whose definition and weights are as follows according to the paper of the ADD 2022 Challenge \cite{yi2022add}. 
$$
WEER = \alpha * EER\_R1 + \beta * EER\_R2
$$
where $\alpha = 0.4$ and $\beta = 0.6$, EER\_R1 and EER\_R2 are the EER of first round and second round evaluation in track 3.2, respectively.
The results obtained on the track 1 adaptation set and the track 3.2 test set of the ADD 2022 dataset confirmed the performance of our score fusion based systems.

\subsection{Results of data augmentation and normalization}
The proposed LCNN model with STFT-1024 features and data augmentation techniques are experimented on the adaptation set of track 1. Table \ref{aug} describes the EER of different data augmentation methods on the track 1 adaptation set. Bare in the table refers to the absence of any data augmentation methods. Kaldi refers to the noise and reverberation addition offline via Kaldi. Online means adding noise and reverberation to the data online in a Kaldi-like manner. Codec means using the codec data mentioned in subsection \ref{dataaug}. Norm means normalizing the audio via sv56.

\begin{table}[ht]
	\centering
	\caption{EER of the STFT-1024-LCNN system with different augmentation methods on the track 1 adaptation set.}
	\label{aug}
	\vspace{2mm}
	\begin{tabular}{ c c }
		\toprule
		\textbf{Augmentation} & \textbf{EER/\%}\\
		\midrule
		Bare & 21.69 \\
		Kaldi & 14.38 \\
		Online & 9.31 \\
		Online \& Norm & 7.38 \\
		Online \& Codec & 7.69 \\
		Online \& Codec \& Norm & 5.61 \\
		\bottomrule
	\end{tabular}
\end{table}

The experimental results show that data augmentation methods with noise and reverberation addition can effectively improve the generalization of the fake audio detection system. In particular, adding reverberation and noise online reduced the EER by 5.07\% compared to the offline method. The reason is that the data is fixed after offline data augmentation, while the randomness of online data augmentation can further enrich the data distribution. Since channel variation is not the main interference in the test set, the codec method bring no significant improvement. The normalization lead to a further decrease in EER by about 2\%.  However, the performance of normalization is not obvious on the test set. On the one hand, this is due to the small amount of adaptation set data. On the other hand, because of the interference such as various noise, the volume difference of the test set data is relatively small. So the performance improvement of the audio normalization is difficult to show on the test set.

\subsection{Results of submitted systems}
Table \ref{result}  presents the performance of the submitted systems for the ADD challenge track 3.2. Fusion2 refers the further integration of partially fake audio detection system in track 2 based on the track 3.2 fusion system above.

\begin{table}[ht]
	\centering
	\caption{EER of final submitted systems for Track 3.2}
	\label{result}
	\vspace{2mm}
	\begin{tabular}{ l c c r}
		\toprule
		\multirow{2}*{\textbf{Feature}} & \multirow{2}*{\textbf{Loss}} & \multicolumn{2}{c}{\textbf{EER\%}} \\
		\cline{3-4}
		& & 1st round & 2nd round \\
		\midrule
		STFT-1024 & AM-Softmax & 10.30 & 14.29 \\
		STFT-1024 & Center Loss & 10.89 & 15.84 \\
		STFT-1024-norm & AM-Softmax & - & 15.75 \\
		STFT-2048 & AM-Softmax & - & 14.11 \\
		STFT-2048 & Center Loss & - & 15.36 \\
		CQT & Center Loss & - & 15.29 \\
		\midrule
		Fusion & - & 9.59 & 13.11 \\
		Fusion2 & - & 10.93 & 11.96 \\
		\bottomrule
	\end{tabular}
\end{table}

With the same features and models, different loss functions have performance differences, as shown by the experimental results. The EER of AM-Softmax Loss based system is about 1\% lower than that of Center Loss.

In terms of data augmentation, while audio normalization can improve performance on the track 1 adaptation set, it causes performance degradation in the second test set of track 3.2 instead.

As far as the fusion system is concerned, the fusion results at the score level bring a relative reduction of 6.9\% and 7.1\% in EER for the first and second rounds respectively, compared to the best single system. Probably due to the fact that the fake audio submitted by other teams in the second round contains fake audio obtained by splicing, the integration with a partially fake audio detection system reduced the EER by 1.15\% in the second round. However, the same system fusion dose not significantly improve the performance in the first round.

\begin{table}[ht]
	\centering
	\caption{EERs of the top-performancing systems in ADD Challenge track 3.2 \cite{yi2022add}. EER\_R1, EER\_R2 and WEER represent the first round of evaluation, the second round of evaluation, and the final weighted EER, respectively.}
	\label{top}
	\begin{tabular}{ l c c r}
		\toprule
		\textbf{ID} & \textbf{EER\_R1} & \textbf{EER\_R2} & \textbf{WEER} \\
		\midrule
		D01\cite{yan2022audio} & 8.6 & 11.1 & 10.1 \\
		D02 & 9.4 & 11.0 & 10.4 \\
		D03 & 8.3 & 12.1 & 10.6 \\
		D04 (Ours) & 9.6 & 12.0 & 11.0 \\
		D05 & 8.6 & 12.8 & 11.1 \\
		\bottomrule
	\end{tabular}
\end{table}

Table \ref{top} shows the EERs of the 5 best performing systems in the ADD Challenge Track 3.2. The gap between the different teams is very small. By fusing the scores of multiple single systems, we obtained competitive results. Since the systems submitted in the first round only fused scores of two systems, and the noise data used in data augmentation is relatively monotonous, the performance was relatively poor. In the second round, benefiting from the fusion of scores with a partially fake audio detection system, the gap between the submitted system and the other teams was further reduced.

\subsection{Score fusion analysis}
Since the performance improvement of score fusion in ADD track3.2 and ASVspoof 2021 is smaller than that of ASVspoof 2019, the reason is worth exploring. The fusion of scores succeeds mainly because sometimes one score misses real samples, but the other score can be used to identify real samples. Explicitly excluding fake samples is not the main mechanism for score fusion works \cite{ulery2006studies}. Therefore, the distribution of scores in the same category from different systems should have a strong correlation.

\begin{figure}[ht]
	\includegraphics[width=\columnwidth]{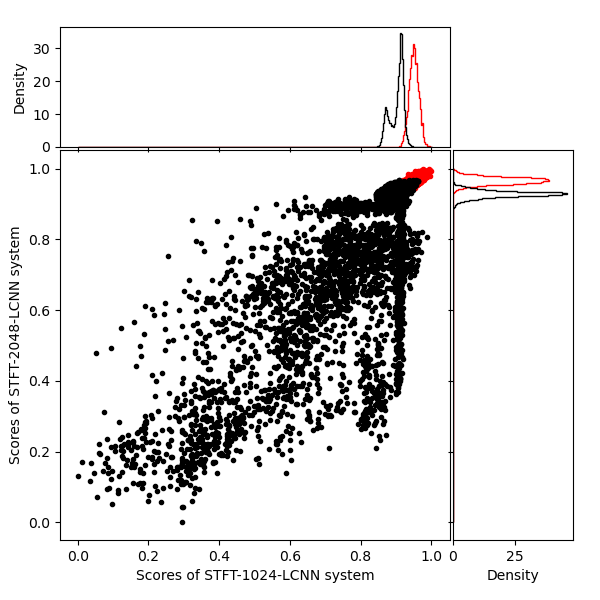}
	\caption{Example of score distributions in ASVspoof 2019 LA, with \textcolor{red}{real in red} and fake in black. The EERs of the two STFT-LCNN-based models are 3.67\% and 3.23\%, respectively. And the EER of equal weighted score fusion is 2.56\%.}
	\Description{Score distribution of two LCNN-based models on ASVspoof 2019 LA. The EERs of the two models are 3.67\% and 3.23\%, respectively. And the EER of the score fusion is 2.56\%, a relative improvement of 21\% compared to the best single-system performance.}
	\label{asvspoof}
\end{figure}

The distribution of scores of the two LCNN-based countermeasures on the ASVspoof 2019 LA evaluation set is shown in the Figure \ref{asvspoof}. Similar to the systems proposed above, the features fed into LCNN models are STFT-1024 and STFT-2048 respectively. The EERs of the two countermeasures on the ASVspoof 2019 LA evaluation partition are 3.67\% and 3.23\%, respectively. By observing the histogram of scores, the difference between the scores of real samples and fake samples is relatively small. And the scores of fake samples have obvious long tail distribution. From the scatter plot of scores, the bonafide samples in red in the figure are highly correlated. The distribution of spoof samples in black, although relatively scattered, there is still a certain correlation. Therefore, the performence of score fusion with an EER of 2.56\% is improved significantly, which is a relative improvement of 21\% compared to the single systems. Although the data labels for the ADD challenge have not been released, the reason for the insufficient fusion can be found by visualizing and comparing the correlation and distribution of scores.

\begin{figure}[ht]
	\includegraphics[width=\columnwidth]{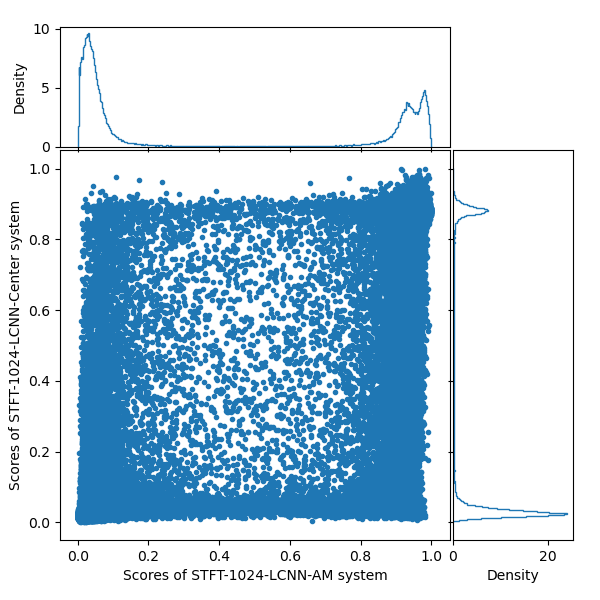}
	\caption{Score distribution and histograms for the two systems in the first round of track 3.2. The axis titles represent the corresponding systems. Where STFT and CQT represent features, and 1024 and 2048 represent FFT bins of features. AM means AM-Softmax Loss and Center means Center Loss.}
	\Description{Score distribution of two LCNN-based models on ADD 2022.}
	\label{round1}
\end{figure}

Figure \ref{round1} shows the score histograms and distribution obtained from the two countermeasures on the test set of ADD Challenge track 3.2 first round. As shown in the histograms, the scores of the test data show a clear polarization. This indicates that the fake audio detection systems are very confident in their judgments. We argue that this is due to the large amount of out-of-distribution data in the test set. Similar conclusions appear in \cite{wang2022estimating}. This "blind confidence" causes different systems to make completely opposite judgments on the same audio data, and the confidence levels are unreasonably high. This leads to the fact that weights are difficult to choose when scores are fused and the weight of one system must dominates. Otherwise the performance of the fusion system will be degraded. 

\begin{figure*}[ht]
	\includegraphics[width=\textwidth]{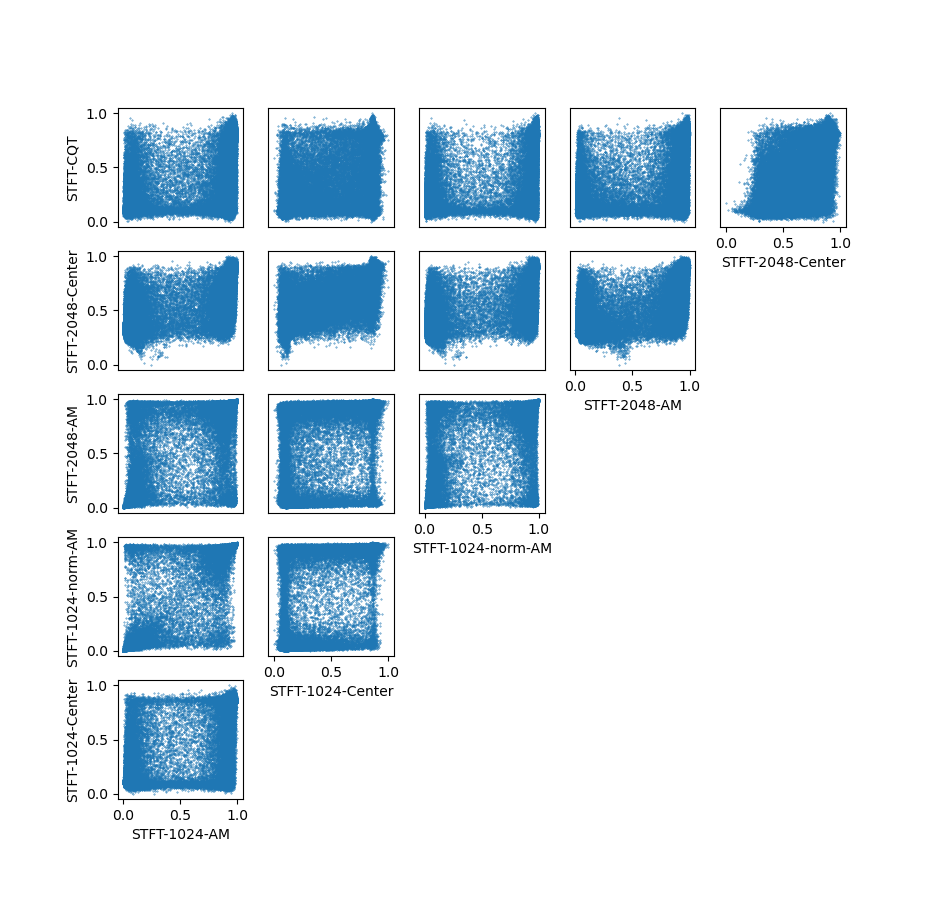}
	\caption{Score distribution between the 6 systems in the second round of track 3.2. The axis titles represent the corresponding systems. Where STFT and CQT represent features, and 1024 and 2048 represent FFT bins of features. AM means AM-Softmax Loss and Center means Center Loss.}
	\label{plot}
\end{figure*}

The scatter plot in Figure \ref{round1} shows the distribution of scores for the two systems in the first round of track3.2. Figure \ref{plot} shows the distributions of scores between the 6 systems in the second round. As shown in figures, The loss function has an effect on the distribution of the data. The scores of the two Center Loss based systems are more closely distributed, but less discriminative. The score distributions of the two AM-Softmax Loss based systems are concentrated toward 0 and 1 in the lower left and upper right corners. In other words, the system scores based on the AM-Softmax Loss are more extreme. Since scores of Center Loss based model is relatively uniform and scores of AM-Softmax Loss based model is extreme, the score distributions of systems based on Center Loss and AM-Softmax Loss tends to more concentrated on two edges. This is due to the different degrees of constraint of the loss function.

Furthermore, for the test dataset of two rounds of the ADD Challenge track 3.2, there is little correlation between the distributions of the scores of all the different systems. Even if the network architectures of these systems are the same, with only differences in features and loss functions, different systems make completely different judgments on whether an audio is fake or not. Combined with the extremes of the scores, it can be inferred that the data distribution of the test set and the training set is significantly different. The system has been unintentionally overfitted to the training set, resulting in a large variation in the detection of the test data by different systems, resulting in an extreme distribution of scores and a lack of correlation..

As a result, the poor performance of system fusion is due to the two reasons above: The system gives scores with high confidence, and the distributions of scores from different systems are poorly correlated. Therefore, even if the scores can get similar EERs, the score fusion is also less effective. The root cause is the overfitting of the model to the training set, while the data in the test set is out of distribution of the training set.

\section{Conclusion}
A deepfake audio detection system submitted to the ADD 2022 challenge track 3.2 is descripted in this paper. The proposed score fusion and LCNN based fake audio detection system achieves a WEER of 11.04\% in track 3.2 final ranking, which is one of the top performing systems. Commonly used noise and reverberation additions can effectively improve the robustness of the system. Improving the offline augmentation to online augmentation can further improve the system performance. The impact of data augmentation, especially the audio normalization on the detection system is also analyzed. Although audio normalization has a significant effect on the adaptation set of track 1, it does not lead to performance improvements in the test sets of track 3.2. This is mainly due to the fact that the test sets are much more complex compared to the adaptation set, with richer volume size variations. The most important thing in the paper is a detailed visual analysis of the reasons for the poor score fusion effect. The score distribution of different systems is visualized and compared with the score distribution of the same system in the ASV spoof 2019 LA where the score fusion effect is obvious. It can be found that the distributions of scores of different systems are extreme and lack correlation due to the over-confidence of the of the judgments made by overfitted system in the out-of-distribution data. This results in no significant effect of score fusion and requires careful selection of weights. Generalization of algorithms in the face of unknown deepfake algorothms and robustness in complex environments remain key issues for fake audio detection systems.

\section{Acknowledgments}
This work is partially supported by the National Key Research and Development Program of China (No. 2021YFC3320103).

\bibliographystyle{ACM-Reference-Format}
\balance
\bibliography{refs}

%
%
%
%
%
%
%
%

\end{document}